\documentclass{article}
\usepackage{times}
\usepackage{amsfonts}
\usepackage{graphicx}
\usepackage[pdfmark]{hyperref}
\begin{document}
\noindent
{\Large  EMERGENT QUANTUM MECHANICS  AS A\\ CLASSICAL, IRREVERSIBLE THERMODYNAMICS}
\vskip1cm
\noindent
{\bf D. Acosta$^{1,a}$, P. Fern\'andez de C\'ordoba$^{2,b}$, J.M. Isidro$^{2,c}$ and J.L.G. Santander$^{3,d}$}\\
${}^{1}$Departamento de Matem\'aticas, Universidad de Pinar del R\'{\i}o,\\ Pinar del R\'{\i}o, Cuba\\
${}^{2}$Instituto Universitario de Matem\'atica Pura y Aplicada,\\ Universidad Polit\'ecnica de Valencia, Valencia 46022, Spain\\
${}^{3}$C\'atedra Energesis de Tecnolog\'{\i}a Interdisciplinar, Universidad Cat\'olica de Valencia,\\ C/ Guillem de Castro 94, Valencia 46003, Spain\\
${}^{a}${\tt dago@mat.upr.edu.cu}, ${}^{b}${\tt pfernandez@mat.upv.es}\\
${}^{c}${\tt joissan@mat.upv.es}, ${}^{d}${\tt martinez.gonzalez@ucv.es} \\
\vskip.5cm
\noindent
{\bf Abstract} We present an explicit correspondence between quantum mechanics and the classical theory of irreversible thermodynamics as developed by Onsager, Prigogine {\it et al}\/.  Our correspondence maps irreversible Gaussian Markov processes into the semiclassical approximation of quantum mechanics. 
Quantum--mechanical propagators are mapped into thermodynamical probability distributions. The Feynman path integral also arises naturally in this setup. The fact that quantum mechanics can be translated into thermodynamical language provides additional support for the conjecture that quantum mechanics is not a fundamental theory but rather an emergent phenomenon, {\it i.e.}, an effective description of some underlying degrees of freedom.\\

\tableofcontents

\section{Introduction}\label{uno}

Emergent physics as a research topic has drawn a lot of attention recently \cite{CARROLL1, HU2}. The very spacetime we live in, as well as the gravitational force that governs it, both appear to be emergent phenomena \cite{HU1, PADDY, VERLINDE}. Quantum mechanics has also been conjectured to be the emergent theory of some underlying deterministic model, in part because of its long--standing conflict with general relativity. There exists a large body of literature on emergent quantum mechanics, some basic references being \cite{ADLER1, THOOFT2, NELSON1}; see also \cite{ADLER2, CARROLL2, CETTO, ELZE1, GROESSING1, THOOFT3, LEE, GROESSING3, NELSON2, STABILE, GROESSING2, SMOLIN} for more recent work. The hypothesis of emergence and the holographic principle \cite{THOOFT1, SUSSKIND} have been hailed as landmarks in the endeavour to arrive at a consistent a theory of quantum gravity.

Without touching on the difficulties facing quantum gravity, a number of interpretational questions and foundational issues arise and remain within a purely quantum--mechanical setup (or, eventually, within a quantum field theory setup, see \cite{THOOFT4}). In this article, following earlier work \cite{NOI}, we will focus on {\it the emergent aspects of quantum mechanics applying a thermodynamical approach}\/. In fact the classical thermodynamics of irreversible processes and fluctuation theory will turn out to share many common features with quantum mechanics---surprisingly, with Feynman's path integral approach to quantum mechanics. Some basic references on the subject of fluctuations and irreversible thermodynamics are \cite{LANDAU, ONSAGERRECIP, ONSAGER, PRIGOGINE1, TISZA}; intriguing questions such as the emergence of macroscopic irreversibility from microscopic reversibility, the arrow of time, and other related puzzles are analysed in \cite{MISRA, PRIGOGINE2}. A more complete list of references can be found in \cite{OLAH}.

Specifically, the purpose of this article is twofold:\\
{\it i)} to establish an explicit correspondence between quantum mechanics on the one hand, and the classical thermodynamics of irreversible processes on the other. We claim validity for this correspondence at least in the Gaussian approximation (which corresponds to the linear response regime in thermodynamics, and to the semiclassical approximation in quantum mechanics); \\
{\it ii)} to use the correspondence just mentioned in order to provide an independent proof of the statement that {\it quantum mechanics is an emergent phenomenon, at least in the semiclassical limit}\/.

With hindsight, once one has realised that quantum mechanics in the Gaussian approximation is a classical thermodynamics in disguise, the emergent nature of quantum theory becomes selfevident---after all, thermodynamics is a paradigm of emergent theories.

\section{The Chapman--Kolmogorov equation in quantum mechanics}\label{dois}

To begin with we present a collection of {\it purely}\/ quantum--mechanical expressions, for which there will be {\it purely}\/ thermodynamical reexpressions using the correspondence we are about to develop.  Although the material of this section is standard, a good general reference is \cite{ZJ}. For simplicity we will restrict to a 1--dimensional configuration space $X$ coordinatised by $x$.

The quantum--mechanical propagator $K\left(x_2,t_2\vert x_1,t_1\right)$ is defined as the amplitude for the conditional probability that a particle starting at $(x_1,t_1)$ end at $(x_2,t_2)$:
\begin{equation}
K\left(x_2,t_2\vert x_1,t_1\right)=\langle x_2\vert  U(t_2-t_1)\vert x_1\rangle, \quad U(t)=\exp\left(-\frac{{\rm i}}{\hbar}t{H}\right).
\label{uniprop}
\end{equation}
Above, $U(t)$ is the unitary time--evolution operator, and $H$ is the quantum Hamiltonian operator. The time--evolution operators satisfy {\it the group property}\/,
\begin{equation}
U(t_1)U(t_2)=U(t_1+t_2),
\label{muchauuu}
\end{equation}
an equation known in statistics already since the 1930's as {\it the Chapman--Kolmogorov equation}\/ \cite{DOOB}. Its solutions satisfy the differential equation
\begin{equation}
{\rm i}\hbar\frac{{\rm d}U}{{\rm d}t}=HU(t),\qquad H={\rm i}\hbar\frac{{\rm d}U}{{\rm d}t}{\Big \vert}_{t=0}.
\label{ouh}
\end{equation}
Using (\ref{uniprop}) we obtain an alternative reexpression of the Chapman--Kolmogorov equation:
\begin{equation}
K\left(x_3,t_3\vert x_1,t_1\right)=\int{\rm d}x_2\,K\left(x_3,t_3\vert x_2,t_2\right)K\left(x_2,t_2\vert x_1,t_1\right).
\label{gruppo}
\end{equation}
Since wavefunctions $\psi$ are {\it unconditional}\/ probability amplitudes, they are related to propagators $K$ (which are {\it conditional}\/ probability amplitudes) as follows:
\begin{equation}
\psi(x_2,t_2)=\int{\rm d}x_1\,K\left(x_2,t_2\vert x_1,t_1\right)\psi(x_1,t_1).
\label{relacao}
\end{equation}
Propagators can be computed via path integrals over configuration space $X$,
\begin{equation}
K\left(x_2,t_2\vert x_1,t_1\right)=\int_{x(t_1)=x_1}^{x(t_2)=x_2}{\rm D}x(t)\,\exp\left\{\frac{{\rm i}}{\hbar}\int_{t_1}^{t_2}{\rm d}t\,L\left[x(t),\dot x(t)\right]\right\},
\label{paz}
\end{equation}
where $L$ is the classical Lagrangian function.
Two simple examples in which the path integral (\ref{paz}) can be evaluated exactly are the free particle and the harmonic potential. For a free particle
we have
\begin{equation}
K^{({\rm free})}\left(x_2,t_2\vert x_1,t_1\right)=\sqrt{\frac{m}{2\pi {\rm i}\hbar\,(t_2-t_1)}}\exp\left[\frac{{\rm i}m}{2\hbar}\frac{(x_2-x_1)^2}{t_2-t_1}\right],
\label{esatto}
\end{equation}
while for a harmonic potential we have, ignoring the caustics,
\begin{equation}
K^{({\rm harmonic})}\left(x_2,t_2\vert x_1,t_1\right)=\sqrt{\frac{m\omega}{2\pi{\rm i}\hbar\sin\left(\omega(t_2-t_1)\right)}}
\label{souri}
\end{equation}
$$
\times \exp\left\{\frac{{\rm i}m}{2\hbar}\frac{\omega}{\sin\left(\omega(t_2-t_1)\right)}\left[(x_2^2+x_1^2)\cos\left(\omega(t_2-t_1)\right)-2x_1x_2\right]\right\}.
$$
When the path integral (\ref{paz}) cannot be computed exactly, an approximate evaluation can still be helpful. For $\hbar\to 0$ we have the semiclassical approximation to the propagator, denoted by $K_{\rm cl}$:
\begin{equation}
K_{\rm cl}\left(x_2,t_2\vert x_1,t_1\right)= Z^{-1}\,\exp\left\{\frac{{\rm i}}{\hbar}\int_{t_1}^{t_2}{\rm d}t\,L\left[x_{\rm cl}(t), \dot x_{\rm cl}(t)\right]\right\},
\label{mezzo}
\end{equation}
where $x_{\rm cl}(t)$ stands for the classical trajectory between $(x_1,t_1)$ and $(x_2,t_2)$, and $Z^{-1}$ is some normalisation factor.\footnote{We will henceforth use the collective notation $Z^{-1}$ to denote all the different normalisation factors that we will not keep track of.}

\section{Fluctuations and irreversible processes}\label{onsamach}

{}For the benefit of the reader, with an eye on later applications, we include below a summary of ref. \cite{ONSAGER}.

\subsection{Thermodynamic forces}\label{tres}

Let a thermodynamical system be given. If we are interested in only a single instant, the probability $P$ of a given state is given by Boltzmann's principle,
\begin{equation}
k_B\ln P=S+ {\rm const},
\label{bolman}
\end{equation}
where $S$ is the entropy of that state. If we are interested in two instants widely separated in time, the probability of given states at each instant is equal to the product of the individual probabilities. A long time lapse makes the states statistically independent. Hence the joint probability of the succession is related to the sum of the two entropies. But if the time lapse is not long, the states will be statistically correlated. It is precisely the laws for irreversible behaviour which tell us the correlations.

Let the thermodynamical state of our system be defined by a set of extensive variables $y^1,\ldots, y^N$. The entropy $S=S(y^1,\ldots, y^N)$ will be a function of all the $y^k$. Its maximum (equilibrium) value will be denoted by $S_0$, and the $y^k$ will be redefined to vanish for the equilibrium state: $S_0=S(0,\ldots, 0)$. The tendency of the system to seek equilibrium is measured by the {\it thermodynamic forces}\/ $Y_k$ defined as
\begin{equation}
Y_k=\frac{\partial S}{\partial y^k}, \quad k=1, \ldots, N.
\label{forza}
\end{equation}
The $Y_k$ are {\it restoring forces}\/ that vanish with the $y^k$.

{}Fluxes are measured by the time derivatives of the $y^k$. The essential physical assumption made here is that {\it irreversible processes are linear,  i.e., they depend linearly on the forces that cause them}\/. Therefore we have\footnote{We use $\tau$ to denote time in the theory of irreversible thermodynamics, and $t$ to denote time in the quantum theory. As will be seen in (\ref{vic}), $\tau$ and $t$ are related by a Wick rotation.}
\begin{equation}
\dot y^i=\frac{{\rm d}y^i}{{\rm d}\tau}=\sum_{j=1}^NL^{ij}\,Y_j, \qquad i=1, \ldots, N.
\label{lineale}
\end{equation}
Onsager's reciprocity theorem states that $L$ is a symmetric matrix \cite{ONSAGERRECIP},
\begin{equation}
L^{ij}=L^{ji}.
\label{reziproko}
\end{equation}
{}Further assuming that $L$ is nonsingular one can solve for the forces in terms of the fluxes:
\begin{equation}
Y_i=\sum_{j=1}^NR_{ij}\dot y^j, \qquad i=1, \ldots, N.
\label{equiserre}
\end{equation}
Thus the rate of production of entropy,
\begin{equation}
\dot S=\sum_{j=1}^N\frac{\partial S}{\partial y^j}\dot y^j=\sum_{j=1}^NY_j\dot y^j,
\label{produk}
\end{equation}
can be expressed in either of two equivalent ways:
\begin{equation}
\dot S=\sum_{i,j=1}^NR_{ij}\dot y^i\dot y^j=\sum_{i,j=1}^NL^{ij}Y_iY_j.
\label{eserrele}
\end{equation}
One defines the {\it dissipation function}\/ $\Phi$ as the following quadratic form in the fluxes:\footnote{We assume $R_{ij}$ to be positive definite. This ensures that $\dot S>0$ as expected of a dissipative process. Indeed, the dissipation function $\Phi$ can be identified with a kinetic energy, $T=\sum_{i,j=1}^Ng_{ij}\dot x^i\dot x^j/2$, where $g_{ij}$ is a certain Riemannian metric on the space spanned by the velocities $\dot x^j$. Identifying $\dot x^j$ with $\dot y^j$ we have $g_{ij}=R_{ij}$.}
\begin{equation}
\Phi:=\frac{1}{2}\sum_{i,j=1}^NR_{ij}\dot y^i\dot y^j.
\label{qadra}
\end{equation}
This function is a potential for the $Y_k$, because $\partial\Phi/\partial\dot y^j=R_{jk}Y_k$. The corresponding quadratic form of the forces,
\begin{equation}
\Psi:=\frac{1}{2}\sum_{i,j=1}^NL^{ij}Y_iY_j,
\label{tik}
\end{equation}
has a similar property, but it should be noticed that it is a function of the {\it state}\/ (since the $Y_k$ depend only on the $y^j$), whereas the numerically equal $\Phi$ is a function of its {\it rate of change}\/.

If we expand the entropy in a Taylor series around equilibrium we have
\begin{equation}
S=S_0-\frac{1}{2}\sum_{i,j=1}^Ns_{ij}y^iy^j+\ldots
\label{teilo}
\end{equation}
The matrix $s_{ij}$ is symmetric and positive definite. Neglect of the higher terms in $y^k$ means the assumption that fluctuations are Gaussian: for Boltzmann's principle (\ref{bolman}) states that the logarithm of the probability of a given fluctuation is proportional to its entropy, or
\begin{equation}
P(y^1,\ldots, y^N)=Z^{-1}\exp\left(\frac{S}{k_B}\right)=Z^{-1}\exp\left(-\frac{1}{2k_B}\sum_{i,j=1}^Ns_{ij}y^iy^j\right).
\label{sarto}
\end{equation}
The assumption of Gaussianity (\ref{teilo}) then implies that the $Y_i$ are linear in the $y^j$:
\begin{equation}
Y_i=-\sum_{j=1}^Ns_{ij}y^j.
\label{llin}
\end{equation}
Thus the phenomenological laws (\ref{equiserre}) become
\begin{equation}
\sum_{j=1}^N\left(R_{ij}\dot y^j+s_{ij}y^j\right)=0.
\label{gus}
\end{equation}

\subsection{Fluctuations}\label{quattro}

Let us now modify the deterministic equations (\ref{equiserre}) to include fluctuations by the addition of a random force $\xi_i$,
\begin{equation}
\sum_{j=1}^NR_{ij}\dot y^j=Y_i+\xi_i,
\label{radon}
\end{equation}
which turns (\ref{equiserre}) into the set of stochastic equations (\ref{radon}). We require that the $\xi_i$ have zero means, which implies that the right--hand side of (\ref{radon}) is a random force with means $Y_i$. For simplicity, as in the quantum--mechanical case, let us set $N=1$, so we have a single variable $y$ obeying the stochastic equation
\begin{equation}
R\dot y+sy=\xi.
\label{unica}
\end{equation}

We will be concerned with the path of $y$ in time under the influence of these random forces. Our aim is to calculate the probability of any path.  For $n$ instants of time $\tau_1<\tau_2<\ldots <\tau_n$ we denote the {\it cumulative distribution function}\/ by $F_n$:
\begin{equation}
F_n\left({y_1\atop \tau_1}{\ldots\atop\ldots}{y_n\atop \tau_n}\right)=P\left(y(\tau_k)\leq y_k,\; k=1,\ldots, n\right).
\label{comol}
\end{equation}
The function $F_n$ tells the probability that the thermodynamical path $y(\tau)$ lie below the barriers $y_1, \ldots, y_n$ at the corresponding instants $\tau_1,\ldots, \tau_n$.  A {\it stationary}\/ process is defined as one whose cumulative distribution function $F_n$ is invariant under arbitrary time shifts $\delta\tau$:
\begin{equation}
F_n\left({y_1\atop \tau_1}{\ldots\atop\ldots}{y_n\atop \tau_n}\right)=
F_n\left({y_1\atop \tau_1+\delta\tau}{\ldots\atop\ldots}{y_n\atop \tau_n+\delta\tau}\right), \qquad\forall\,\delta\tau\in\mathbb{R}.
\label{arby}
\end{equation}
Physically this describes an {\it aged}\/ system, one that has been left alone long enough that any initial conditions have worn off, or been forgotten. Thus we consider entropy creation as a loss of information: a dissipative system forgets its past.

Alongside $F_n$,  the {\it probability density function}\/ $f_n$ is defined such that the product
\begin{equation}
f_n\left({y_1\ldots y_n}\atop{\tau_1\ldots \tau_n}\right){\rm d}y_1\cdots{\rm d}y_n
\label{pedeefe}
\end{equation}
gives the probability that a thermodynamical path pass through gates of width ${\rm d}y_k$.

We will also be interested in conditional probabilities.  The {\it conditional probability function}\/ for the $(n+1)$th event given the previous $n$,
\begin{equation}
F_1\left({y_{n+1}\atop \tau_{n+1}}{\Big\vert}{y_1\atop \tau_1}{\ldots\atop\ldots}{y_n\atop \tau_n}\right)=P\left(y(\tau_{n+1})=y_{n+1}\;{\Big \vert}\;y(\tau_k)=y_k,\; k=1,\ldots, n\right),
\label{caillou}
\end{equation}
is defined implicitly as follows:
\begin{equation}
F_{n+1}\left({y_1\atop \tau_1}{\ldots\atop\ldots}{y_{n+1}\atop \tau_{n+1}}\right)
\label{muchasf}
\end{equation}
$$
=\int_{-\infty}^{y_1}{\rm d}\tilde y_1\cdots\int_{-\infty}^{y_n}{\rm d}\tilde y_n\,F_1\left({y_{n+1}\atop \tau_{n+1}}{\Big \vert}{\tilde y_1\atop \tau_1}{\ldots\atop\ldots}{\tilde y_n\atop \tau_n}\right)\,{\rm d}F_{n}\left({\tilde y_1\atop \tau_1}{\ldots\atop\ldots}{\tilde y_n\atop \tau_{n}}\right).
$$
Correspondingly, the {\it conditional probability density function}\/ $f_1$ is defined such that
\begin{equation}
f_1\left({y_k\atop \tau_k}{\Big \vert}{y_{k-1}\atop \tau_{k-1}}\right){\rm d}y_k\,{\rm d}y_{k-1}
\label{unamas}
\end{equation}
equals the probability that a thermodynamical path pass through a gate of width ${\rm d}y_k$ at time $\tau_k$, {\it given}\/ that it passed through a gate of width ${\rm d}y_{k-1}$ at time $\tau_{k-1}$.

\subsection{Markov processes}\label{cinque}

A Markov process is defined as one whose conditional probabilities are independent of all but the immediately preceding instant \cite{DOOB}:
\begin{equation}
F_1\left({y_{n+1}\atop \tau_{n+1}}{\Big \vert}{y_{1}\atop \tau_1}{\ldots\atop\ldots}{y_{n}\atop \tau_n}\right)=
F_1\left({y_{n+1}\atop \tau_{n+1}}{\Big\vert}{y_{n}\atop \tau_n}\right).
\label{macof}
\end{equation}
Intuitively: a Markov system has a short memory. For a Markov process (\ref{muchasf}) and (\ref{macof}) imply
\begin{equation}
f_n\left({y_{1}\ldots y_{n}}\atop{\tau_1\ldots \tau_n}\right)=f_1\left({y_{n}\atop \tau_n}{\Big\vert}{y_{n-1}\atop \tau_{n-1}}\right)
\cdots f_1\left({y_{2}\atop \tau_{2}}{\Big \vert}{y_{1}\atop \tau_{1}}\right)\, f_1\left({y_{1}\atop \tau_1}\right).
\label{danno}
\end{equation}
Now $f_1\left({y_{1}\atop \tau_1}\right)$ is known from Boltzmann's principle (\ref{bolman}). Hence, by stationarity, all that is needed in order to obtain the distribution function for an arbitrary number of gates is to evaluate the conditional probability density function
\begin{equation}
f_1\left({y_{2}\atop \tau + \delta\tau}{\Big\vert}{y_{1}\atop \tau}\right),
\label{pdf}
\end{equation}
which depends only on $\delta\tau$, being independent of $\tau$. Thus the $n$--gate problem reduces to the 2--gate problem.

\subsection{Gaussian processes}\label{seis}

A Gaussian stochastic process is one whose probability density function is a Gaussian distribution. Let us set, in (\ref{unica}),
\begin{equation}
\gamma:=\frac{s}{R}.
\label{fed}
\end{equation}
Then the conditional probability function for a Gaussian process is given by \cite{ONSAGER}
\begin{equation}
f_1\left({y_{2}\atop \tau+\delta\tau}{\Big\vert}{y_{1}\atop \tau}\right)=\frac{1}{\sqrt{2\pi}}\frac{s/k_B}{\sqrt{1-{\rm e}^{-2\gamma\delta\tau}}}
\exp\left[-\frac{s}{2k_B}\frac{\left(y_{2}-{\rm e}^{-\gamma\delta\tau}y_{1}\right)^2}{1-{\rm e}^{-2\gamma\delta\tau}}\right].
\label{prpa}
\end{equation}
Now eqn. (\ref{prpa}), together with (\ref{danno}), constitutes the solution to the problem of finding the probability of any path in a Gaussian Markov process. We also remark that (\ref{prpa}) correctly reduces to the one--gate distribution function (\ref{sarto}) for $\delta\tau\to\infty$.

Next let us divide the interval $(\tau,\tau+\delta\tau)$ into $n$ equal subintervals of length $\delta\tau/n$:
\begin{equation}
\tau_1=\tau,\quad \tau_2=\tau_1+\frac{\delta\tau}{n},\quad \ldots, \quad \tau_{n+1}=\tau+\delta\tau.
\label{divizio}
\end{equation}
Then we have
\begin{equation}
f_1\left({y_{n+1}\atop \tau_{n+1}}{\Big\vert}{y_1\atop \tau_1}\right)=\int{\rm d}y_n\cdots\int{\rm d}y_2\, f_1\left({y_{n+1}\atop \tau_{n+1}}{\Big\vert}{y_n\atop \tau_n}\right)\cdots f_1\left({y_{2}\atop \tau_{2}}{\Big\vert}{y_1\atop \tau_1}\right).
\label{muul}
\end{equation}
This is again the Chapman--Kolmogorov equation. The integral above extends over all the $n-1$ intermediate gates. Using (\ref{muul}) one can reexpress (\ref{prpa}) in the following alternative form \cite{ONSAGER}:
\begin{equation}
f_1\left({y_{n+1}\atop \tau_{n+1}}{\Big\vert}{y_1\atop \tau_1}\right)=Z^{-1}\exp\left\{-\frac{1}{4k_B}\int_{\tau_1}^{\tau_{n+1}}{\rm d}\tau\,R\left[\dot y(\tau)+\gamma y(\tau)\right]^2\right\}_{\rm min},
\label{piszminint}
\end{equation}
subject to $y(\tau_1)=y_1$, $y(\tau_{n+1})=y_{n+1}$. The subscript {\it min}\/ refers to the fact that argument of the exponential is to be evaluated along the trajectory that minimises the integral.

The one--gate distribution is obtained from the conditional distribution $f_1\left({y_2\atop\tau_2}{\Big\vert}{y_1\atop\tau_1}\right)$ by taking $\tau_1=-\infty$ and $y_1=0$ (because the aged system certainly was at equilibrium long ago). Thus we set $n=1$ in (\ref{piszminint}) and define the {\it thermodynamical Lagrangian function}\/ ${\cal L}$ as
\begin{equation}
{\cal L}\left[\dot y(\tau), y(\tau)\right]:=\frac{R}{2}\left[\dot y(\tau)+\gamma y(\tau)\right]^2.
\label{agraz}
\end{equation}
The dimension of ${\cal L}$ is entropy per unit time, instead of energy. However, our map between mechanics and thermodynamics will justify the denomination ``Lagrangian".  The Euler--Lagrange equation for a minimum value of the integral in (\ref{piszminint}) is
\begin{equation}
\ddot y-\gamma^2y=0.
\label{moto}
\end{equation}
The solution to the above that satisfies the boundary conditions $y(\tau=-\infty)=0$ and $y(\tau=\tau_2)=y_2$ is
\begin{equation}
y(\tau)=y_2{\rm e}^{\gamma(\tau-\tau_2)}.
\label{disi}
\end{equation}
Evaluating the integral in (\ref{piszminint}) along this extremal trajectory leads to
\begin{equation}
f_1\left({y_2\atop\tau_2}{\Big\vert}{0\atop -\infty}\right)=f_1\left(y_2\atop \tau_2\right)=Z^{-1}\,\exp\left[-\frac{s}{2k_B}(y_2)^2\right].
\label{konss}
\end{equation}
This result is in agreement with what one expects from Boltzmann's principle (\ref{bolman}) in the Gaussian approximation (\ref{teilo}).

{}Finally substituting (\ref{konss}) into (\ref{muul}), we obtain the thermodynamical analogue of the quantum--mechanical relation (\ref{relacao}):
\begin{equation}
f_1\left(y_2\atop \tau_2\right)=\int{\rm d}y_1\,f_1\left({y_2\atop\tau_2}{\Big\vert}{y_1\atop\tau_1}\right)f_1\left(y_1\atop \tau_1\right).
\label{tambien}
\end{equation}
This concludes our summary of ref. \cite{ONSAGER}.

\section{The map between quantum mechanics and irreversible thermodynamics}\label{siete}

The Wick rotation
\begin{equation}
\tau={\rm i}t
\label{vic}
\end{equation}
between the thermodynamical evolution parameter $\tau$ and the quantum--mechanical time variable $t$ is the first entry in our dictionary between classical irreversible thermodynamics and quantum mechanics.

\subsection{Path integrals in irreversible thermodynamics}\label{nou}

The concept of a path integral can be traced back to the Chapman--Kolmogorov equation. Indeed letting $n\to\infty$ in (\ref{divizio}) and using (\ref{muul}), the right--hand side of  (\ref{piszminint}) becomes a path integral {\it over the thermodynamical configuration space}\/ $Y$:
\begin{equation}
f_1\left({y_{2}\atop \tau_{2}}{\Big\vert}{y_1\atop \tau_1}\right)=\int_{y(\tau_1)=y_1}^{y(\tau_2)=y_2}{\rm D}y(\tau)\,
\exp\left\{-\frac{1}{4k_B}\int_{\tau_1}^{\tau_{2}}{\rm d}\tau\,R\left[\dot y(\tau)+\gamma y(\tau)\right]^2\right\}.
\label{cepe}
\end{equation}
Thus it turns out that (\ref{piszminint}) actually equals the semiclassical approximation (as per (\ref{mezzo})) to the path integral (\ref{cepe}). This latter  expression for the distribution function $f_1$ in terms of a path integral is implicit in ref. \cite{ONSAGER}---but actually never written down explicitly in that paper; see however \cite{GRAHAM}.

Dropping in (\ref{agraz}) the term proportional to $\dot yy$ (a total derivative), we redefine the thermodynamical Lagrangian function ${\cal L}$ to be
\begin{equation}
{\cal L}\left[\dot y(\tau), y(\tau)\right]=\frac{R}{2}\left[\dot y^2(\tau)+\gamma^2y^2(\tau)\right].
\label{agra}
\end{equation}
We observe that $\dot y^2(\tau)$ and $y^2(\tau)$ in ${\cal L}$ carry the same relative sign.  Similarly dropping in (\ref{cepe}) the term proportional to $\dot yy$, we can rewrite the path integral using (\ref{agra}) as
\begin{equation}
f_1\left({y_{2}\atop \tau_{2}}{\Big\vert}{y_1\atop \tau_1}\right)=\int_{y(\tau_1)=y_1}^{y(\tau_2)=y_2}{\rm D}y(\tau)\,
\exp\left\{-\frac{1}{2k_B}\int_{\tau_1}^{\tau_{2}}{\rm d}\tau\,{\cal L}\left[\dot y(\tau), y(\tau)\right]\right\}.
\label{cepele}
\end{equation}
The path integral (\ref{cepele}) is the thermodynamical analogue of the path integral (\ref{paz}) that defines the quantum--mechanical propagator. Thus setting $n=1$ in (\ref{piszminint}), dropping the total derivative $\dot yy$, and replacing the integrand with the thermodynamical Lagrangian (\ref{agra}) leads to the Gaussian approximation to (\ref{cepele}):
\begin{equation}
f_1\left({y_{2}\atop \tau_{2}}{\Big\vert}{y_1\atop \tau_1}\right)=Z^{-1}\,\exp\left\{-\frac{1}{2k_B}\int_{\tau_1}^{\tau_{2}}{\rm d}\tau\,{\cal L}\left[\dot y_{\rm cl}(\tau), y_{\rm cl}(\tau)\right]\right\}.
\label{nouw}
\end{equation}
Here ${\cal L}\left[\dot y_{\rm cl}(\tau), y_{\rm cl}(\tau)\right]$ stands for the evaluation of (\ref{agra}) along the classical trajectory $y_{\rm cl}(\tau)$
that satisfies the equations of motion (\ref{moto}). In this way (\ref{nouw}) is seen to correspond to the semiclassical approximation for the quantum--mechanical propagator, given in (\ref{mezzo}). On the thermodynamical side, the quantum--mechanical semiclassical approximation translates as the assumption of Gaussianity for the stochastic forces $\xi$ and for the entropy $S$, as well as the assumption of linearity between forces and fluxes (which leads up to the quadratic forms (\ref{qadra}) and (\ref{tik})).

\subsection{Propagators from thermodynamical distributions}\label{huit}

The next entry in our dictionary relates quantum--mechanical wavefunctions and propagators to thermodynamical distribution functions. Within the Gaussian approximation we use throughout, this entry will refer to the free particle and the harmonic oscillator. We first we need to identify certain mechanical variables with their thermodynamical partners. Specifically, we will make the following replacements:\footnote{A dimensionful conversion factor must be understood as implicitly contained in the replacement $x\leftrightarrow y$, whenever needed.}
\begin{equation}
\omega\leftrightarrow\gamma,\qquad \frac{m\omega}{\hbar}\leftrightarrow\frac{s}{2k_B}, \qquad x\leftrightarrow y.
\label{bazz}
\end{equation}

To begin with, one expects the squared modulus of the wavefunction $\vert\psi\vert^2$ to be related to the 1--gate distribution function $f_1\left({y\atop \tau}\right)$, while the propagator $K$ must correspond to a 2--gate distribution function $f_1\left({y_{2}\atop\tau_2}{\vert}{y_{1}\atop \tau_1}\right)$.  Indeed the 1--gate distribution function (\ref{konss}) gives the squared modulus of the ground state $\psi_0(x)=\exp\left(-m\omega x^2/2\hbar\right)$ of the harmonic oscillator once the replacements (\ref{vic}), (\ref{bazz}) are applied:
\begin{equation}
f_1\left({x\atop {\rm i}t}\right)=Z^{-1}\,\exp\left(-\frac{m\omega}{\hbar}x^2\right)=\vert\psi_{0}^{\rm (harmonic)}(x)\vert^2.
\label{kirch}
\end{equation}
With the appropriate choices for the constants $m$ and $\omega$, (\ref{kirch}) can also represent a free wavepacket. Next we turn to propagators $K$. Elementary algebra brings the conditional probability function for a Gaussian process (\ref{prpa}) into the form
\begin{equation}
f_1\left({y_{2}\atop\tau}{\Big\vert}{y_{1}\atop 0}\right)=\frac{s}{2k_B}\frac{{\rm e}^{\gamma\tau/2}}
{\sqrt{\pi\,\sinh\left(\gamma\tau\right)}}
\exp\left[-\frac{s}{2k_B}\frac{\left({\rm e}^{\gamma\tau/2}y_{2}-{\rm e}^{-\gamma\tau/2}y_{1}\right)^2}{2\,\sinh\left(\gamma\tau\right)}\right].
\label{prpazz}
\end{equation}
We will also be interested in the limit $\gamma\to 0$ of the above:
\begin{equation}
f_1\left({y_{2}\atop \tau}{\Big\vert}{y_{1}\atop 0}\right)_{\gamma\to 0}\simeq\frac{s}{2k_B}\frac{1}{\sqrt{\pi\,\gamma\tau}}
\exp\left[-\frac{s}{2k_B}\frac{(y_2-y_1)^2}{2\gamma\tau}\right].
\label{gamusin}
\end{equation}
Using (\ref{vic}) and (\ref{bazz}), the free quantum--mechanical propagator (\ref{esatto}) follows from (\ref{gamusin}):
\begin{equation}
K^{\rm (free)}(x_2,t\vert x_1,0)=\sqrt{\frac{k_B}{s}}\,f_1\left({x_2\atop {\rm i}t}{\Big\vert} {x_1\atop 0}\right)_{\gamma\to 0}.
\label{speck}
\end{equation}
The case when $\gamma$ is nonvanishing requires some more work. Again (\ref{vic}) and (\ref{bazz}) allow one to relate the conditional probability (\ref{prpazz}) to the harmonic propagator (\ref{souri}) as follows:
\begin{equation}
f_1\left({x_2\atop {\rm i}t}{\Big\vert}{x_1\atop 0}\right)=\exp\left(\frac{{\rm i}\omega t}{2}-\frac{\Delta V}{\hbar\omega}\right)\sqrt{\frac{2m\omega}{\hbar}}\,K^{\rm (harmonic)}\left(x_2,t\vert x_1,0\right),
\label{novita}
\end{equation}
where $V(x)=kx^2/2$ is the harmonic potential and $\Delta V=V(x_2)-V(x_1)$. As had to be the case, (\ref{novita}) correctly reduces to (\ref{speck}) when $\omega\to 0$.
The square roots present in (\ref{speck}) and (\ref{novita}) ensure that these two equations are dimensionally correct.

\subsection{Integrability {\it vs.}\/ square--integrability}\label{intvssqint}

Under our correspondence, the squared modulus of the wavefunction $\vert\psi\vert^2$ gets mapped into the {\it unconditional}\/ probability density $f_1\left({y_{1}\atop \tau_1}\right)$, while the propagator $K$ gets mapped into the {\it conditional}\/ probability density $f_1\left({y_{2}\atop \tau_{2}}\vert {y_{1}\atop \tau_{1}}\right)$. One should bear in mind, however, that the quantum--mechanical objects $\psi, K$ are probability {\it amplitudes}\/, while the thermodynamical objects $f_1$ are true probabilities. Therefore quantum mechanics is not just the Wick rotation of classical, irreversible thermodynamics---it is also the {\it square root}\/ thereof, so to speak, because of the Born rule. In order to address this question in mode detail we need to recall some background mathematics; see ref. \cite{THIRRING} for a physics--oriented approach, and also \cite{BAEZ} for a recent discussion of some of the issues analysed later in this section.

Let $M$ be a measure space, and denote by $L^p(M)$ the Banach space\footnote{The space $L^p(M)$ is complex or real according to whether its elements $f$ are taken to be complex--valued or real--valued functions on $M$. For quantum--mechanical applications we will consider the complex case, while thermodynamical applications require the real case. For generality, this summary assumes all spaces complex.}
\begin{equation}
L^p(M)=\left\{f:M\rightarrow\mathbb{C},\;\vert\vert f\vert\vert_{p}<\infty\right\},\quad
\vert\vert f\vert\vert_{p}:=\left(\int_M\vert f\vert^p\right)^{1/p}, \quad 0<p<\infty.
\label{banajuno}
\end{equation}
It turns out that $L^p(M)$ is a Hilbert space only when $p=2$. Moreover, $L^p(M)$ and $L^q(M)$ are linear duals of each other whenever $1/p+1/q = 1$. Two particular cases of this duality will interest us. The first one is $p=2, q=2$, the other one is $p=1, q=\infty$.

When $p=2$ we have that $L^2(M)$ is selfdual, the duality being given by the scalar product: $\langle\cdot\vert\cdot\rangle :L^{2}(M)\times L^2(M)\longrightarrow\mathbb{C}$. The corresponding algebra of bounded operators is ${\cal L}(L^2(M))$, a noncommutative $C^*$--algebra with respect to operator multiplication. Complex conjugation in ${\cal L}(L^2(M))$ consists in taking the adjoint operator, while the noncommutativity is that of matrix multiplication.

The operator algebra ${\cal L}(L^p(M))$ is also a Banach algebra for any $p>0$, and not just for $p=2$. However, only when $p=2$ is a ${\cal L}(L^p(M))$ a $C^*$--algebra, because only when $p=2$ does ${\cal L}(L^p(M))$ possess a complex conjugation.

Set now $p=1$. The dual of $L^1(M)$ is $L^{\infty}(M)$. Elements of the latter are measurable, essentially bounded functions $f$ with a finite norm $\vert\vert f\vert\vert_{\infty}$:
\begin{equation}
L^{\infty}(M)=\left\{f:M\rightarrow\mathbb{C},\;\vert\vert f\vert\vert_{\infty}<\infty\right\},\qquad \vert\vert f\vert\vert_{\infty}
:={\rm sup}_{z\in M}\{\vert f(z)\vert\}.
\label{banajdos}
\end{equation}
The duality between $L^1(M)$ and  $L^{\infty}(M)$ is
\begin{equation}
(\cdot\vert\cdot ):L^{\infty}(M)\times L^1(M)\longrightarrow\mathbb{C}, \qquad (f\vert\rho):=\int_Mf\rho,
\label{dwall}
\end{equation}
for any $f\in L^{\infty}(M)$ and any $\rho\in L^{1}(M)$. Now $L^{\infty}(M)$ also qualifies as a $C^*$--algebra, the multiplication law being pointwise multiplication of functions (hence commutative), and the complex conjugation being that of the functions $f$. An important difference with respect to the previous case is that ${\cal L}(L^2(M))$ is noncommutative, whereas $L^{\infty}(M)$ is commutative.

We will henceforth write $X$ for the space $M$ when dealing with the mechanical configuration space, and $Y$ when referring to the thermodynamical configuration space.

Textbook quantum mechanics regards quantum states as unit rays within $L^2(X)$, while physical observables ${\cal O}$ are represented by selfadjoint operators ${\cal O}\in{\cal L}(L^2(X))$.\footnote{We ignore the mathematical subtleties due to the fact that ${\cal O}$ is generally an unbounded operator, hence generally not an element of ${\cal L}(L^2(X))$, because this fact is immaterial to the discussion.}  On the other hand, the natural framework for the theory of irreversible thermodynamics is the {\it real}\/ Banach space $L^1(Y)$ and its dual, the {\it real}\/ Banach algebra $L^{\infty}(Y)$. Thermodynamical states are probability distributions $\rho\in L^1(Y)$, that is, {\it real}\/ functions, normalised as per $\int_Y\rho=1$. Thermodynamical observables are {\it real}\/ functions $f\in L^{\infty}(Y)$. Thus $\int_Yf\rho$ in (\ref{dwall}) equals the average value of the physical quantity $f$ in the state described by $\rho$.

Clearly the thermodynamical setup is not quite as sophisticated as its mechanical counterpart. As opposed to the {\it complex}\/ Hilbert space $L^2(X)$, the {\it real}\/ Banach space $L^1(Y)$ does not know about the existence of the imaginary unit $i$. In the absence of a complex conjugation to implement time reversal, the thermodynamical setup necessarily describes {\it irreversible}\/ processes. Moreover, there exists no scalar product on $L^1(Y)$. Correspondingly there is no notion of a selfadjoint operator in ${\cal L}(L^1(Y))$---in fact, thermodynamical observables are elements of a very different space, $L^{\infty}(Y)$.\footnote{In particular, the {\it real}\/ space $L^{\infty}(Y)$ is a Banach algebra but not a $C^*$--algebra.}

The previous differences notwithstanding, we can establish a map between quantum--mechanical states/observables and their thermodynamical counterparts, as we do next. We treat observables first, and discuss states later.

It is reasonable to identify real thermodynamical averages $(f\vert\rho)$ with quantum mechanical expectation values $\langle\psi\vert{\cal O}\vert\psi\rangle$ of selfadjoint operators ${\cal O}$, something like
\begin{equation}
\int_Yf\rho=(f\vert\rho)\leftrightarrow\langle\psi\vert{\cal O}\vert\psi\rangle=\int_X\psi^*{\cal O}\psi,
\label{kori}
\end{equation}
where the correspondence denoted by $\leftrightarrow$ has yet to be given a precise meaning. For this we can assume diagonalising ${\cal O}$ by a (complete, orthonormal) set of eigenstates $\psi_i\in L^2(X)$, so we can replace the right--hand side of (\ref{kori}) with the corresponding eigenvalue $\lambda_i$. We want to define a functional $f$ for the left--hand side of (\ref{kori}). A sensible definition actually involves a collection of constant functionals $f_i$, each one of them equal to the corresponding eigenvalue $\lambda_i$:
\begin{equation}
f_i:Y\longrightarrow\mathbb{R},\qquad f_i(y)=\lambda_i, \qquad \forall y\in Y.
\label{koriez}
\end{equation}
Since the eigenvalues $\lambda_i$ are constants and the density $\rho$ can be normalised to unity, the imprecise correspondence (\ref{kori}) can be replaced with the precise dictionary entry
\begin{equation}
\int_Yf_i\rho=(f_i\vert\rho)=\lambda_i=\langle\psi_i\vert{\cal O}\vert\psi_i\rangle=\int_X\psi^*{\cal O}\psi.
\label{koripp}
\end{equation}
This generalises in the obvious way to the case of a set of commuting observables ${\cal O}_k$. Noncommuting observables, not being simultaneously diagonalisable, lead to the impossibility of simultaneously defining the corresponding thermodynamical functionals $f$ on the left--hand side of (\ref{koripp}). We will examine the thermodynamical analogue of quantum commutators in a forthcoming publication.

So much for the observables; now we turn to the states. Since thermodynamical probabilities are elements of $L^1(Y)$ while quantum--mechanical amplitudes belong to $L^2(X)$, we would like to define some map of $L^2(X)$ into $L^1(Y)$, or viceversa. Given $\psi\in L^2(X)$, one's first instinct is to set $\rho:=\vert\psi\vert^2$ because then $\rho\in L^1(X)$; this is of course the Born rule. The attentive reader will have noticed that we actually need $\rho\in L^1(Y)$: it is generally meaningless to equate $\rho$ to $\vert\psi\vert^2$---or to any other function of $\psi$, for that matter. We will proceed ahead under the simplifying assumption that $X=Y$.

The usual Born map $b$ is defined as
\begin{equation}
b:L^2(X)\longrightarrow L^1(X),\quad b(\psi):=\vert\psi\vert^2.
\label{inyecbor}
\end{equation}
This map is obviously not 1--to--1, so it fails to be an injection. As such it possesses no inverse. We will however use the formal notation $b^{-1}$ to denote the map
\begin{equation}
b^{-1}:L^1(X)\longrightarrow L^2(X),\quad b^{-1}(\rho):=\sqrt{\rho}\,{\rm e}^{\frac{{\rm i}}{\hbar}\varphi},
\label{jotabor}
\end{equation}
where $\varphi$ is taken as the solution to the continuity equation
\begin{equation}
\dot\rho+\nabla\cdot\left(\rho\nabla\varphi\right)=0
\label{malung}
\end{equation}
that is well known from the Madelung transformation. Moreover, if $b^{-1}(\rho)$ satisfies the Schroedinger equation, then $\varphi$ must of course equal the action integral $I=\int{\rm d}t\,L$, and thus satisfy the {\it quantum}\/ Hamilton--Jacobi equation \cite{MATONE}. Although the map $b^{-1}$ also fails to be an injection, we use the notation $b^{-1}$ because $bb^{-1}(\rho)=\rho$. Aside from this difficulty about the lack of injectivity, $b$ and $b^{-1}$ provide us with the required maps from quantum--mechanical states into thermodynamical distribution functions, and viceversa.

The Chapman--Kolmogorov equation (\ref{muul}), written below for $n=2$,
\begin{equation}
f_1\left({y_{3}\atop \tau_{3}}{\Big\vert}{y_1\atop \tau_1}\right)=\int{\rm d}y_2\, f_1\left({y_{3}\atop \tau_{3}}{\Big\vert}{y_2\atop \tau_2}\right)f_1\left({y_{2}\atop \tau_{2}}{\Big\vert}{y_1\atop \tau_1}\right),
\label{jjaevz}
\end{equation}
is the thermodynamical analogue of the quantum--mechanical equation (\ref{gruppo}). This leads us to the following point. Our correspondence maps $f_1\left({y_2\atop\tau_2}{\Big\vert}{y_1\atop\tau_1}\right)$, which is a conditional probability, into $K(x_2,t_2\vert x_1,t_1)$, which is an {\it amplitude}\/ for a conditional probability.
In other words, under our correspondence, the Born rule does {\it not}\/ apply to the map between conditional probabilities, although it does apply to the map between unconditional probabilities. There is nothing wrong with this. Indeed, $f_1$ and $K$ satisfy the respective Chapman--Kolmogorov equations (\ref{jjaevz}) and (\ref{gruppo}). Regarding the latter as matrix equations (which is what they are), they read formally $f_1\times f_1 = f_1$ and $K\times K = K$. That is, squaring $f_1$ and $K$ as matrices (which is how they should be squared, since $f_1$ and $K$ are operators), they are idempotent. It therefore makes sense {\it not}\/ to impose the Born rule on the map between $K$ and $f_1$.

\subsection{Entropy {\it vs.}\/ action}\label{thebonrl}

To complete our dictionary between quantum mechanics and irreversible thermodynamics we postulate the following correspondence between the action integral $I$ and the entropy $S$:
\begin{equation}
\mbox{(mechanics)}\quad\frac{{\rm i}}{\hbar}I\leftrightarrow\frac{1}{k_B}S\quad\mbox{(thermodynamics)},
\label{korrespondenzia}
\end{equation}
up to a numerical, dimensionless factor. Now the Wick rotation (\ref{vic}) replaces ${\rm i}I$ with the Euclidean action $I_E$, so we could just as well write
\begin{equation}
\mbox{(mechanics)}\quad\frac{1}{\hbar}I_E\leftrightarrow\frac{1}{k_B}S\quad\mbox{(thermodynamics)},
\label{sinflecha}
\end{equation}
again up to a numerical, dimensionless factor. We observe that both $I$ and $S$ independently satisfy an extremum principle. We also note that the respective fluctuation theories\footnote{These fluctutations are of course measured with respect to the corresponding mean values of $I$ and $S$ as given by their extremals.} in the Gaussian approximation are obtained upon taking the exponential. Thus exponentiating (\ref{korrespondenzia}) we arrive at the wavefunction
\begin{equation}
\psi=\sqrt{\rho}\,\exp\left(\frac{{\rm i}}{\hbar}I\right)
\label{uvekabe}
\end{equation}
and at the Boltzmann distribution function (\ref{bolman}):
\begin{equation}
\rho_B=Z^{-1}\exp\left(\frac{1}{k_B}S\right).
\label{masbolman}
\end{equation}
We should point out that the correspondence (\ref{korrespondenzia}), (\ref{sinflecha}) has also been found to hold in independent contexts, long ago by de Broglie \cite{BROGLIE} and more recently {\it e.g.} in \cite{NOI, BANERJEE}.

Applying the Born rule we set the Boltzmann probability density $\rho_B$ equal to the quantum--mechanical probability density $\vert\psi\vert^2$:
\begin{equation}
\rho_B=\vert\psi\vert^2=\rho.
\label{egalite}
\end{equation}
(See ref. \cite{BACCIAGALUPPI} for distributions other than the {\it squared}\/ modulus). Hence
\begin{equation}
\rho=Z^{-1}\exp\left(\frac{1}{k_B}S\right).
\label{zustand}
\end{equation}
Substitution of (\ref{zustand}) into (\ref{uvekabe}) yields an elegant expression for the wavefunction
\begin{equation}
\psi=Z^{-1/2}\exp\left(\frac{1}{2k_B}S\right)\exp\left(\frac{{\rm i}}{\hbar}I\right),
\label{beautex}
\end{equation}
combining thermodynamics and quantum mechanics into a single formula.

Implicitly assumed in (\ref{beautex}) is the identification of mechanical variables $x$ and thermodynamical variables $y$, as already done in (\ref{bazz}). One can now define the {\it complex--valued action}\/ ${\cal I}(x)$\footnote{While the entropy $S$ is a true function of $x$, the action integral $I$ is actually a {\it functional}\/ of $x(t)$. However, in (\ref{complejax}) we need $I$ within the exponential defining $\psi$. To this end, $I$ is to be evaluated along {\it the}\/ classical trajectory starting at a certain given point and ending at a variable endpoint $x$. This amounts to regarding $I$ as a true function of $x$ and no longer as a functional.}
\begin{equation}
{\cal I}(x):=\frac{1}{2k_B}S+\frac{{\rm i}}{\hbar}I.
\label{complejax}
\end{equation}
in order to write
\begin{equation}
\psi(x)=Z^{-1/2}\exp\left({\cal I}(x)\right)
\label{bellax}
\end{equation}
as the semiclassical wavefunction (\ref{beautex}), where
\begin{equation}
Z=\int{\rm d}x\,\vert\exp\left({\cal I}(x)\right)\vert^2.
\label{summex}
\end{equation}
We realise that  the correspondence (\ref{korrespondenzia}), (\ref{sinflecha}) leads naturally to the existence of a complexified action such as (\ref{complejax}), which expresses {\it a fundamental symmetry between entropy and mechanical action}\/.

{}Finally we would like to point out that complexified action functionals have also been considered recently in ref. \cite{NIELSEN}.

\section{Discussion}\label{dotze}

We can summarise this article in the following statements:\\
{\it i)} we have succeeded in formulating a correspondence between standard quantum mechanics, on the one hand, and the classical thermodynamics of irreversible processes, on the other;\\
{\it ii)} this correspondence holds at least in the Gaussian approximation (the latter being defined in quantum mechanics as the semiclassical limit, and in thermodynamics as the regime of linearity between forces and fluxes);\\
{\it iii)} this possibility of encoding of quantum--mechanical information in thermodynamical terms provides an independent proof of the statement that quantum mechanics is an an emergent phenomenon.

Specifically, our correspondence between semiclassical quantum mechanics and Gaussian irreversible thermodynamics includes the following points of section \ref{siete}:\\
{\it i)} we have shown that the path--integral representation for quantum--mechanical propagators is already present in the thermodynamical description of classical dissipative phenomena (section \ref{nou});\\
{\it ii)} we have mapped thermodynamical distribution functions into quantum--mechanical propagators (section \ref{huit});\\
{\it iii)} we have constructed an explicit correspondence between quantum--mechanical states and thermodynamical states, and also an analogous correspondence between quantum--mechanical observables and thermodynamical observables (section \ref{intvssqint});\\
{\it iv)} we have grounded our correspondence in the existence of a fundamental symmetry between mechanical action and entropy (section \ref{thebonrl}).\\
In order to make this paper selfcontained we have also included, in section \ref{onsamach}, a crash course in classical irreversible thermodynamics, the latter considered in the linear approximation. Presumably, the theory of irreversible thermodynamics beyond the linear regime should allow one to extend the present correspondence beyond the semiclassical approximation of quantum mechanics. 

Having mapped {\it quantum}\/ mechanics into {\it classical}\/ irreversible thermodynamics raises another old question, {\it viz.}\/, the issue of how sharply, how univocally defined is the divide between {\it quantumness}\/ and {\it classicality}\/. This issue has also been addressed, from the viewpoint of emergent theories, in ref. \cite{ELZE2}; we defer our own contribution to the subject until a forthcoming publication. However we would like to briefly touch upon the emergence property of {\it spacetime}\/---not from a gravitational perspective, but from a purely quantum--mechanical viewpoint. If spacetime is an emergent phenomenon, as widely conjectured, then everything that makes use of spacetime concepts must necessarily be emergent, too. Quantum mechanics is no exception, unless one succeeds in constructing a quantum--mechanical formalism that is entirely free of spacetime notions. Progress towards this latter goal has been achieved along lines based on noncommutative geometry (see \cite{SARDA} and references therein). A more modest approach is to try and directly map quantum mechanics into thermodynamics, as done here and elsewhere. It turns out that spacetime arises as an emergent concept {\it also}\/ in our quantum--mechanical approach, if only because our correspondence has required replacing space variables $x$ with thermodynamical variables $y$. Thus, indirectly, we have also furnished (admittedly cirmcumstantial) evidence of the emergence property of spacetime.

It was Einstein's dream to see quantum mechanics formulated as an ensemble theory in which uncertainties would {\it not}\/ have a fundamental ontological status.  Instead, Einstein would have uncertainties and fluctuations arise as a consequence of {\it the statistical nature}\/ of the description of an underlying {\it deterministic}\/ system (see \cite{KHRENNIKOV, NIEUWENHUIZEN} and refs. therein). Thermodynamical fluctuation theory thus appears to be the archetypal example that Einstein would presumably have liked for quantum mechanics to be modelled upon.

Actually it has been known since the early days of quantum mechanics that the (free) Schroedinger equation can be interpreted as the standard heat equation in imaginary time, so the thermodynamical connection has always existed. An unavoidable consequence of imaginary time is that real (decaying) exponentials replace imaginary (oscillatory) exponentials. This is the hallmark of dissipation. Thus quantum mechanics can be thought of as a dissipative phenomenon that becomes conservative only in stationary states \cite{VITIELLO1, VITIELLO2, THOOFT2}---that little $i$ in the Schroedinger equation makes a big difference \cite{KAUFFMAN}.\\

After completion of this work we became aware of ref. \cite{RUUGE}, where topics partially overlapping with those treated here are discussed.\\

\noindent
{\bf Acknowledgements}  J.M.I. would like to thank the organisers of the Heinz von Foerster Congress on Emergent Quantum Mechanics (Vienna, Austria, Nov. 2011) for stimulating a congenial atmosphere of scientific exchange, and for the interesting discussions that followed.\\
{\it  Willst Du erkennen? Lerne zu handeln!---Heinz von Foerster}\/.


\begin{thebibliography}{99}

\bibitem{NOI}
D. Acosta, P. Fern\'andez de C\'ordoba, J.M. Isidro and J.L.G. Santander, {\it An Entropic Picture of Emergent Quantum Mechanics}, Int. J. Geom. Meth. Mod. Phys. {\bf 9} (2012) 1250048, {\tt arXiv:1107.1898 [hep-th]}.

\bibitem{ADLER1}
S. Adler, {\it Quantum Theory as an Emergent Phenomenon}, Cambridge University Press, Cambridge (2004).

\bibitem{ADLER2}
S. Adler, {\it Quantum Theory as an Emergent Phenomenon: Foundations and Phenomenology},  J. Phys. Conf. Ser. {\bf 361} (2012) 012002.

\bibitem{BACCIAGALUPPI}
G. Bacciagaluppi, {\it Non--Equilibrium in Stochastic Mechanics}, J. Phys. Conf. Ser. {\bf 361} (2012) 012017.

\bibitem{BAEZ}
J. Baez and B. Fong, {\it A Noether Theorem for Markov Processes}, {\tt arXiv:1203.2035 [math-ph]}.

\bibitem{BANERJEE}
R. Banerjee, {\it From Black Holes to Emergent Gravity}, Int. J. Mod. Phys. {\bf D19} (2010) 2365, {\tt arXiv:1005.4286 [gr-qc]}.

\bibitem{VITIELLO1}
M. Blasone, P. Jizba and G. Vitiello, {\it Dissipation and Quantization}, Phys. Lett. {\bf A287} (2001) 205, {\tt arXiv:hep-th/0007138}.

\bibitem{VITIELLO2}
M. Blasone, P. Jizba and G. Vitiello, {\it Dissipation, Emergent Quantization, and Quantum Fluctuations}, in {\it Decoherence and Entropy in Complex Systems, Selected Lectures from DICE 2002}, H.-T. Elze (ed.), Lecture Notes in Physics {\bf 633}, Springer, Berlin (2004).

\bibitem{BROGLIE}
L. de Broglie, {\it La Thermodynamique de la Particule Isol\'ee}, Gauthier--Villars, Paris (1964).

\bibitem{CARROLL1}
R. Carroll, {\it On the Emergence Theme of Physics}, World Scientific, Singapore (2010).

\bibitem{CARROLL2}
R. Carroll, {\it Remarks on Osmosis, Quantum Mechanics, and Gravity}, J. Phys. Conf. Ser. {\bf 361} (2012) 012010,  {\tt arXiv:1104.0383 [gr-qc]}.

\bibitem{CETTO}
A. Cetto, L. de la Pe$\tilde{\rm n}$a and A. Vald\'es--Hern\'andez, {\it Quantization as an Emergent Phenomenon Due to Matter--Zeropoint Field Interaction},
J. Phys. Conf. Ser. {\bf 361} (2012) 012013.

\bibitem{DOOB}
J. Doob, {\it Stochastic Processes}, Wiley, New York (1953).

\bibitem{ELZE1}
H.-T. Elze, {\it The Attractor and the Quantum States},  Int. J. Qu. Info. {\bf 7} (2009) 83, {\tt arXiv:0806.3408 [quant-ph]}.

\bibitem{ELZE2}
H.-T. Elze, {\it Linear Dynamics of Quantum--Classical Hybrids}, Phys. Rev. {\bf A85} (2012) 052109, {\tt arXiv:1111.2276 [quant-ph]};\\
{\it Four Questions for Quantum--Classical Hybrid Theory}, J. Phys. Conf. Ser. {\bf 361} (2012) 012004, {\tt arXiv:1202.3448 [quant-ph]}.

\bibitem{MATONE}
A. Faraggi and M. Matone, {\it The Equivalence Postulate of Quantum Mechanics: Main Theorems}, {\tt arXiv:0912.1225 [hep-th]}.

\bibitem{SARDA}
G. Giachetta, L. Mangiarotti and G. Sardanashvily, {\it Geometric and Algebraic Topological Methods in Quantum Mechanics}, World Scientific, Singapore (2005).

\bibitem{GRAHAM}
R. Graham, {\it Path Integral Formulation of General Diffusion Processes}, Z. Phys. {\bf B26} (1977) 281.

\bibitem{GROESSING1}
G. Groessing, S. Fussy, J. Mesa Pascasio and H. Schwabl, {\it The Quantum as an Emergent System}, J. Phys. Conf. Ser. {\bf 361} (2012) ) 012008, {\tt arXiv:1205.3393 [quant-ph]}.

\bibitem{THOOFT1}
G. 't Hooft, {\it Dimensional Reduction in Quantum Gravity}, {\tt arXiv:gr-qc/\\9310026}.

\bibitem{THOOFT2}
G. 't Hooft, {\it Emergent Quantum Mechanics and Emergent Symmetries}, AIP Conf. Proc. {\bf 957} (2007) 154, {\tt arXiv:0707.4568 [hep-th]}.

\bibitem{THOOFT3}
G. 't Hooft, {\it Quantum Mechanics from Classical Logic}, J. Phys. Conf. Ser. {\bf 361} (2012) 012024;\\
{\it Relating the Quantum Mechanics of Discrete Systems to Standard Canonical Quantum Mechanics}, {\tt arXiv:1204.4926 [quant-ph]}.

\bibitem{THOOFT4}
G. 't Hooft, {\it Duality between a Deterministic Cellular Automaton and a Bosonic Quantum Field Theory in $1+1$ Dimensions}, {\tt arXiv:1205.4107 [quant-ph]}.

\bibitem{HU1}
B. Hu, {\it Gravity and Nonequilibrium Thermodynamics of Classical Matter}, Int. J. Mod. Phys. {\bf D20} (2011) 697, {\tt arXiv:1010.5837 [gr-qc]}.

\bibitem{HU2}
B. Hu, {\it Emergence: Key Physical Issues for Deeper Philosphical Enquiries},  J. Phys. Conf. Ser. {\bf 361} (2012) 012003, {\tt arXiv:1204.1077 [physics.hist-ph]}.

\bibitem{KAUFFMAN}
L. Kauffman, {\it Eigenforms, Discrete Processes and Quantum Processes}, J. Phys. Conf. Ser. {\bf 361} (2012) 012034, {\tt arXiv:1109.1892 [math-ph]}.

\bibitem{KHRENNIKOV}
A. Khrennikov, {\it  ``Einstein's Dream" -- Quantum Mechanics as Theory of Classical Random Fields}, {\tt arXiv:1204.5172 [quant-ph]}.

\bibitem{LANDAU}
L. Landau and E. Lifschitz, {\it Statistical Physics, Part 1}\/, vol. 5 of {\it Course of Theoretical Physics}\/, Pergamon Press, Oxford (1980).

\bibitem{LEE}
J.-W. Lee, {\it Quantum Mechanics Emerges from Information Theory Applied to Causal Horizons}, Found. Phys. {\bf 41} (2011) 744, {\tt arXiv:1005.2739 [hep-th]}.

\bibitem{GROESSING3}
J. Mesa Pascasio, S. Fussy, H. Schwabl and G. Groessing, {\it Classical Simulation of Double Slit Interference via Ballistic Diffusion}, J. Phys. Conf. Ser. {\bf 361} (2012) 012041, {\tt  arXiv:1205.4521 [quant-ph]}.

\bibitem{MISRA}
B. Misra, I. Prigogine and M. Courbage, {\it From Deterministic Dynamics to Probabilistic Descriptions}, Proc. Natl. Acad. Sci. USA {\bf 76} (1979) 3607;\\
{\it Lyapounov Variable: Entropy and Measurement in Quantum Mechanics}, Proc. Natl. Acad. Sci. USA {\bf 76} (1979) 4768.

\bibitem{NIELSEN}
K. Nagao and H. Nielsen, {\it Formulation of Complex Action Theory}, {\tt arXiv:1104.3381 [quant-ph]};\\
{\it Theory Including Future not Excluded---Formulation of Complex Action Theory II---}, {\tt arXiv:1205.3706 [quant-ph]}.

\bibitem{NELSON1}
P. Nelson, {\it Derivation of the Schroedinger Equation from Newtonian Mechanics}, Phys. Rev. {\bf 150} (1966) 1079.

\bibitem{NELSON2}
P. Nelson, {\it Review of Stochastic Mechanics}, J. Phys Conf. Ser. {\bf 361} (2012) 012011.

\bibitem{NIEUWENHUIZEN}
T.  Nieuwenhuizen, {\it Towards Einstein's Dream of a Unified Field Theory: Reports from a Journey on a Long and Winding Road},
J. Phys. Conf. Ser. {\bf 361} (2012) 012036.

\bibitem{OLAH}
N. Olah, {\it Einsteins Trojanisches Pferd: eine Thermodynamische Deutung der Quantentheorie}, Springer, Wien (2011).

\bibitem{ONSAGERRECIP}
L. Onsager, {\it Reciprocal Relations in Irreversible Processes. I.}, Phys. Rev. {\bf 37} (1931) 405;\\
{\it Reciprocal Relations in Irreversible Processes. II.}, Phys. Rev. {\bf 38} (1931) 2265.

\bibitem{ONSAGER}
L. Onsager and S. Machlup, {\it Fluctuations and Irreversible Processes}, Phys. Rev. {\bf 91} (1953) 1505;\\
{\it Fluctuations and Irreversible Processes. II. Systems with Kinetic Energy}, Phys. Rev. {\bf 91} (1953) 1512.

\bibitem{PADDY}
T. Padmanabhan, {\it Lessons from Classical Gravity about the Quantum Structure of Spacetime}, J. Phys. Conf. Ser. {\bf 306} (2011) 012001, {\tt arXiv:1012.4476 [gr-qc]}.

\bibitem{PRIGOGINE1}
I. Prigogine, {\it Introduction to Thermodynamics of Irreversible Processes}, Interscience, New York (1961);\\
{\it Time, Structure and Fluctuations}, Nobel Prize Lecture (1977).

\bibitem{PRIGOGINE2}
I. Prigogine and Cl. George, {\it The Second Law as a Selection Principle: The Microscopic Theory of Dissipative Processes in Quantum Systems}, Proc. Natl. Acad. Sci. USA {\bf 80} (1983) 4590.

\bibitem{RUUGE}
A. Ruuge, {\it Pauli Problem in Thermodynamics}, {\tt arXiv:1208.2919 [math-ph]}.

\bibitem{STABILE}
M. Sakellariadou, A. Stabile and G. Vitiello, {\it Noncommutative Spectral Geometry, Dissipation and the Origin of Quantization},
J. Phys. Conf. Ser. {\bf 361} (2012) 012025.

\bibitem{GROESSING2}
H. Schwabl, J. Mesa Pascasio, S. Fussy and G. Groessing, {\it Quantum Features Derived from the Classical Model of a Bouncer--Walker Coupled to a Zero--Point Field},
J. Phys. Conf. Ser. {\bf 361} (2012) 012021, {\tt arXiv:1205.4519 [quant-ph]}.

\bibitem{SMOLIN}
L. Smolin, {\it  Could Quantum Mechanics be an Approximation to Another Theory?}, {\tt arXiv:quant-ph/0609109}.

\bibitem{SUSSKIND}
L. Susskind, {\it The World as a Hologram}, J. Math. Phys. {\bf 36} (1995) 6377, {\tt arXiv:hep-th/9409089}.

\bibitem{THIRRING}
W. Thirring, {\it Quantum Mathematical Physics}, 2nd edition, Springer, Berlin (2002).

\bibitem{TISZA}
L. Tisza and I. Manning, {\it Fluctuations and Irreversible Thermodynamics}, Phys. Rev. {\bf 105} (1957) 1695.

\bibitem{VERLINDE}
E. Verlinde, {\it On the Origin of Gravity and the Laws of Newton}, JHEP {\bf 1104} (2011) 029,  {\tt arXiv:1001.0785[hep-th]}.

\bibitem{ZJ}
J. Zinn--Justin, {\it Path Integrals in Quantum Mechanics}, Oxford University Press, Oxford (2005).


\end{thebibliography}
\end{document}